\documentclass[epsfig,12pt]{article}
\usepackage{graphicx}
\usepackage{epsfig}

\title{The value of $B_K$ from the experimental data on CP-violation in
 $K$-mesons and up-to-date values of CKM matrix parameters.}
\author{E.A.Andriyash \footnote{andriash@heron.itep.ru}, \\
{\small Moscow State University and ITEP, Russia} \\
G.G.Ovanesyan \footnote{ovanesyn@heron.itep.ru}, \\
{\small Moscow Institute of Physics and Technologies and ITEP, Russia} \\
M.I.Vysotsky \footnote{vysotsky@heron.itep.ru},\\
{\small ITEP, Russia.}
}
\date{}
\begin{document}

\maketitle

\begin{abstract}
The difference between induced by box diagram quantity $\tilde
\epsilon$ and experimentally measured value of $\epsilon$ is
determined and used to obtain the value of $\tilde \epsilon$ with
high precision. Present day knowledge of CKM matrix elements
(including B-factory data), allows us to obtain from the Standard
Model expression for $\tilde \epsilon$ the value of parameter
$B_K$: $B_K = 0.89 \pm0.16$. It turns out to be very close to the
result of vacuum insertion, $B_K = 1$.
\end{abstract}

\newpage


\newpage

\section{Introduction.}

It is well known that CP - violation in $K^0 - \bar K^0$ mixing is
described by the parameter $\tilde \epsilon$. Within the SM, this
parameter is given by box diagrams. It depends in particular on
the CKM matrix elements, to which vertices of box diagrams are
proportional. On the other hand, the experimentally measured
parameters are $\epsilon$ and $\epsilon'$. $\epsilon$ and
$\epsilon'$ enter the measured ratios of decay amplitudes of kaons
into $\pi \pi$ states. These amplitudes are superpositions of
amplitudes $A(K^0 \to (\pi \pi)_{I}) = A_I e^{i \delta_I}$ of kaon
decays into states with definite isospin $I=0,2$, $A_I$ are weak
amplitudes, $\delta_I$ are strong rescattering phases of
$\pi$-mesons. The parameter $\epsilon$ can be expressed as
\cite{Wolf}:

\begin{eqnarray}\label{1}
\epsilon = \tilde\epsilon + i \frac{Im A_0}{Re A_0}.
\end{eqnarray}

Within the SM and in the standard parametrization of CKM matrix,
$Im A_0$ originates from the so-called strong penguin diagrams.
Amplitude $A_2$ also has an imaginary part which originates from
electro-weak penguin diagrams. That is why $Im A_0
>> Im A_2$.

Taking into account that the phases of $\epsilon$ and $\tilde
\epsilon$ are approximately $\frac{\pi}{4}$ \cite{Wolf}, from
Eq.(\ref{1}) we obtain:

\begin{equation}\label{2}
|\tilde \epsilon| \approx |\epsilon| - \frac{1}{\sqrt{2}} \frac{Im
A_0}{Re A_0}.
\end{equation}

The estimation of $\displaystyle{\frac{Im A_0}{Re A_0}}$ was done
in \cite{paper}. This term appears to be a $5-9\%$ correction to
the value of $\tilde \epsilon$ in Eq.(\ref{2}). Provided that we
have estimated the right-hand side of Eq.(\ref{2}) with the help
of Eq.(\ref{tilde epsilon}) we can determine parameter $B_K$,
which parameterizes hadronic matrix element.
Of course, for this purpose we need to know the
values of CKM matrix elements that enter Eq.(\ref{tilde epsilon}).

The parameters $\bar \rho$ and $\bar \eta$ of CKM matrix appear to
be constrained without using the value of $\tilde \epsilon$ in the
fit. Thus we perform the fit of CKM matrix parameters without
using constraint from $\tilde \epsilon$ in it. Then we determine
$B_K$ from Eqs.(\ref{2}),(\ref{tilde epsilon}). Our result is $B_K
= 0.89 \pm0.16$.

This result is close to the result of vacuum insertion: $B_K = 1$.
As discussed in \cite{Ioffe}, the insertions of $\pi$-mesons
states should be taken into account. These insertions form a
sign-alternating series, who's terms depend on the cutoff momentum
of $\pi$-mesons. This cutoff can be reasonably chosen to be $200 -
500 \ MeV$  (at larger virtualities $\pi$-mesons do not exist).
Then the sign-alternating series converges quickly, and one
can take only first two terms. Thus taking into account the
insertions of $\pi$-mesons states lowers $B_K$, and the agreement
with our result improves further.

The lattice result of $B_K$ calculation is $B_K = 0.87 \pm 0.06
\pm 0.14_{quench}$ \cite{Lel}. We see that our result is very close
to it.

The paper is organized as follows: in Section~\ref{teps} we
discuss various estimations of
the value of $\displaystyle{\frac{Im A_0}{Re A_0}}$.
In Section~\ref{fit} we perform the fit of CKM
matrix parameters without using constraint from $\tilde\epsilon$
in it. In Section~\ref{extraction} we determine $B_K$ and compare
it with other results of calculation of $B_K$. Finally, we make
our conclusion in Section~\ref{concl}.

\section{Estimation of the numerical value of $\epsilon - \tilde \epsilon$.}\label{teps}

In this section we review the estimation of $\epsilon - \tilde
\epsilon$ \cite{paper}. We discuss the following three methods.
First, one can use the experimental data on CP-violation in
semileptonic $K_L$-decays, namely parameter $\delta_L$. This
method possesses large uncertainty and also at the level of two
sigmas contradicts the experimental value of
$\displaystyle{\frac{\epsilon'}{\epsilon}}$. Second, one can
obtain the lower bound on $\epsilon - \tilde \epsilon$ from the
experimental value of $\displaystyle{\frac{\epsilon'}{\epsilon}}$.
This lower bound is important in understanding the relative
magnitude of the second term in Eq.(\ref{2}), it turns out to be
$\ge 5\%$. Third, we use the results of direct computation of $Im
A_0$ in the ratio $\displaystyle{\frac{Im A_0}{Re A_0}}$,
substituting the experimental value of $Re A_0$. This gives us a
reliable estimate of $\displaystyle{\frac{Im A_0}{Re A_0}}$ with
moderate error,which we use in the bulk of the paper.

First, we estimate the value of $\tilde \epsilon$ from the
experimental results on CP-violation in semileptonic $K_L$ decays:

\begin{eqnarray}
&&\delta_L = \frac{\Gamma (K_L \rightarrow l^{+} \nu \pi^{-}) -
\Gamma (K_L \rightarrow l^{-} \bar \nu \pi^{+})}{\Gamma (K_L
\rightarrow l^{+} \nu \pi^{-}) + \Gamma (K_L \rightarrow l^{-}
\bar \nu \pi^{+})} \approx 2 Re \tilde \epsilon. \nonumber\\
&& |\tilde \epsilon| =\frac{\delta_L}{2 \cos \phi } ,
\end{eqnarray}

where $\phi = arg (\tilde \epsilon)$.

Now let us substitute the experimental data. For $\phi$ we use
$\phi = (43.50\pm 0.05)^\circ $ \cite{PDG}. World average value of
$\delta_L$, published in \cite{KTev}, contains new KTev result:
$\delta_L = (3.307 \pm 0.063)\times 10^{-3}$. Na48 collaboration
recently obtained: $\delta_L = (3.317 \pm 0.100)\times 10^{-3}$
\cite{Na48}. Averaging these two numbers we get: $\delta_L = (3.310
\pm 0.053)  \times 10^{-3}$. This leads to the following value of
$\tilde\epsilon$:

\begin{equation}
|\tilde \epsilon| = (2.282 \pm 0.037)\times 10^{-3}.
\end{equation}

From Eq.(\ref{2}) with the help of  Eq.(\ref{5}) we can find the
corresponding value of
$\displaystyle{\frac{Im A_0}{Re A_0}}$:

\begin{equation}\label{105}
\frac{Im A_0}{Re A_0} = (0.03 \pm 0.56)\times 10^{-4}.
\end{equation}

We will show below, that this number almost contradicts the
present experimental value of
$\displaystyle{\frac{\epsilon'}{\epsilon} = (1.67 \pm 0.26)\times
10^{-3}}$ \cite{PDG}.

Second method of estimation of $\displaystyle{\frac{Im A_0}{Re
A_0}}$, which gives the lower bound on it, uses the experimental
value of $\displaystyle{\frac{\epsilon'}{\epsilon}}$. The
expression for $\displaystyle{\frac{\epsilon'}{\epsilon}}$ is
usually presented as follows \cite{Wolf}:

\begin{equation}\label{3}
\frac{\epsilon'}{\epsilon} = \frac{i}{\sqrt 2} e^{i(\delta_2
-\delta_0)} \frac{1}{\epsilon} \left[ \frac{Im A_2}{Re A_0} - w
\frac{Im A_0}{Re A_0} \right]~,
\end{equation}

Let us neglect the term proportional to $Im A_2$ in Eq.(\ref{3}),
which comes from the EW penguins. Taking
into account that $(\delta_0 - \delta_2)_{exp} = 42 \pm 4^{o}$
\cite{Chell}, we obtain the following expression for
$\displaystyle{\frac{Im A_0}{Re A_0}}$ from Eq.(\ref{3}):

\begin{equation}\label{4}
\frac{Im A_0}{Re A_0} \approx - \frac{\sqrt 2 |\epsilon|}{w}
\frac{\epsilon'}{\epsilon}.
\end{equation}

Substituting experimental values from \cite{PDG}, we get:
\begin{eqnarray}\label{5}
&& \frac{\epsilon'}{\epsilon} = (1.67 \pm 0.26) \times 10^{-3},
\quad w = 0.045, \quad |\epsilon| = 2.284(14) \times 10^{-3}
\Longrightarrow \nonumber\\ && \frac{Im A_0}{Re A_0} = - (1.2 \pm
0.2) \times 10^{-4}.
\end{eqnarray}

In this way we get the following value of $|\tilde\epsilon|$:

\begin{equation}\label{6}
|\tilde \epsilon| = 2.37(2) \times 10^{-3}.
\end{equation}

Since, according to Eq.(\ref{3}), the contribution of EW penguins
partially cancels that of QCD penguin, the value
$|\displaystyle{\frac{Im A_0}{Re A_0}}| = (1.2 \pm 0.2) \times
10^{-4}$ should be considered as a lower bound on
$|\displaystyle{\frac{Im A_0}{Re A_0}}|$ and Eq.(\ref{6}) is a
lower bound on $|\tilde \epsilon|$. Thus the central value of
$\displaystyle{\frac{Im A_0}{Re A_0}}$, obtained from semileptonic
$K_L$-decays, Eq.(\ref{105}), almost contradicts the experimental
value of $\displaystyle{\frac{\epsilon'}{\epsilon}}$.

Finally, the reliable way to estimate $\displaystyle{\frac{Im
A_0}{Re A_0}}$, result of which we will use in the next sections is
to use the experimental value of $Re A_0$ and theoretical
value for $Im A_0$.

Calculation of $Re A_0$ and $Im A_0$, as well as $Re A_2$ and $Im
A_2$, has a long history. The calculation of $Re A_0$ and $Re
A_2$ was performed in order to explain the $\Delta I =
\frac{1}{2}$ rule in kaon decays and the calculation of $Im A_0$ and
$Im A_2$ - in
order to explain the observed value of
$\displaystyle{\frac{\epsilon'}{\epsilon}}$.

In this paper we perform the calculation of $Im A_0$ to the
following accuracy: the Wilson coefficient is calculated to LO
and hadronic matrix element is calculated in naive factorization
approximation. The details are presented in Appendix, and here we
only quote the result:

\begin{eqnarray}\label{65}
\frac{Im A_0}{Re A_0} = -( 3.2^{+1.1}_{-0.8} )\times 10^{-4}.
\end{eqnarray}

We note that the results of computation of $Im A_0$ (see
\cite{Gamiz} - \cite{Jamin} and refs. therein) performed by a
large number of people lie in the same ballpark.

Finally, from Eq.(\ref{65}) we can determine the value of $\tilde
\epsilon$:

\begin{equation}\label{10}
|\tilde \epsilon| = (2.51 \pm 0.07) \times 10^{-3}.
\end{equation}

This number is our final result, and we will use it in
Section~\ref{extraction}.

\section{Fit of the parameters of CKM matrix }\label{fit}

We use in our fit of the CKM matrix experimentally measured values
of modulus of matrix elements
$V_{ud}$,$V_{us}$,$V_{ub}$,$V_{cd}$,$V_{cs}$, $V_{cb}$ and also
$sin 2\alpha$, $sin 2\beta$, $sin 2\gamma$ and $\Delta m_{B_d}$.
Note that we do not use $\tilde \epsilon$ in fit, since we plan to
determine the value of $B_K$ with the help of the fit results.

We assume these experimentally measured data to be normally
distributed. Also the theoretical uncertainties are treated as
normally distributed. Let us note that other people treat
theoretical uncertainties in other way \cite{CKMfitter},
\cite{UTFit}.

The table of input parameters looks like:

\begin{tabular}{|c|c|c|}
  \hline
  Parameter & Value & Standard Deviation \\
  \hline
  $|V_{ud}| \cite{PDG}$ & 0.9738 & 0.0005 \\
  $|V_{us}| \cite{PDG}$ & 0.2200 & 0.0026 \\
  $|V_{ub}| \cite{PDG}$ & 0.00367 & 0.00047 \\
  $|V_{cd}| \cite{PDG}$ & 0.224 & 0.012 \\
  $|V_{cs}| \cite{PDG}$ & 0.996 & 0.013 \\
  $|V_{cb}| \cite{PDG}$ & 0.0413 & 0.0015 \\
  $sin 2\alpha$ \cite{alpha} & -0.21 & 0.46 \\
  $sin 2\beta$ \cite{PDG} & 0.736 & 0.049 \\
  $sin 2\gamma$ \cite{gamma}& 0.69 & 0.58 \\
  \hline
\end{tabular}

The $\chi^2$ expression which we minimize looks like:
\begin{eqnarray}\label{chisq}
&&\chi^2(A,\lambda,\bar\rho,\bar\eta)=\left(\frac{V^{theo}_{ud}-V^{exp}_{ud}}{\sigma_{V_{ud}}}\right)^2+
  \left(\frac{V^{theo}_{us}-V^{exp}_{us}}{\sigma_{V_{us}}}\right)^2+
  \left(\frac{V^{theo}_{ub}-V^{exp}_{ub}}{\sigma_{V_{ub}}}\right)^2
+\left(\frac{V^{theo}_{cd}-V^{exp}_{cd}}{\sigma_{V_{cd}}}\right)^2+\nonumber\\
&&+\left(\frac{V^{theo}_{cs}-V^{exp}_{cs}}{\sigma_{V_{cs}}}\right)^2+
  \left(\frac{V^{theo}_{cb}-V^{exp}_{cb}}{\sigma_{V_{cb}}}\right)^2
  +\left(\frac{\Delta m^{theo}_{B_d}-\Delta
m^{exp}_{B_d}}{\sigma_{\Delta_m}}\right)^2+
  \left(\frac{sin 2\alpha^{theo}-sin 2\alpha^{exp}}{\sigma_{sin
2\alpha}}\right)^2+\nonumber\\
&&  +\left(\frac{sin 2\beta^{theo}-sin
2\beta^{exp}}{\sigma_{sin2\beta}}\right)^2+
  \left(\frac{sin 2\gamma^{theo}-sin 2\gamma^{exp}}{\sigma_{sin
2\gamma}}\right)^2 ,
\end{eqnarray} where theoretical expressions
depend on four Wolfenstein parameters: $A$, $\lambda$, $\bar \rho$ and
$\bar \eta$. Expression (\ref{chisq}) was minimized varying them.

Here are our results: $$ \lambda = 0.224 \pm 0.002 \qquad \qquad
\alpha^{[deg]}=100 \pm 5 $$
$$ A = 0.82 \pm 0.03 \qquad \qquad \beta^{[deg]} = 23 \pm 2$$
$$\bar\rho = 0.22 \pm 0.04 \qquad \qquad \gamma^{[deg]} =57 \pm 5
 $$ $$ \bar\eta =
0.34 \pm 0.02 \qquad \qquad \hspace{48pt}$$  $$ \chi^2/n.d.o.f. =
8.1/5 \;\; .$$

For comparison, we present the results of the fit, made by
CKMfitter Group \cite{CKMfitter} and UTfit Collaboration
\cite{UTFit}:

\begin{tabular}{|c|c|c|}
  \hline
   & CKMfitter & UTfit \\
  \hline
  $\lambda  $&  $0.226 \pm 0.002$ & $0.226 \pm 0.002$ \\
  $A$ &$ 0.80^{+0.03}_{-0.02}$ &  \\
  $\bar \rho$ & $0.19^{+0.09}_{-0.07}$ & $0.17\pm 0.05$ \\
  $\bar \eta$ & $0.36^{+0.05}_{-0.04}$ & $0.35 \pm 0.03$ \\
  \hline
\end{tabular}

\vspace{12pt}

\begin{tabular}{|c|c|c|}
  \hline
   & CKMfitter & UTfit \\
  \hline
  $\alpha^{[deg]}  $&  $94^{+12}_{-10}$ & $94 \pm 8$ \\
  $\beta^{[deg]} $ &$ 23.8^{+2.1}_{-2.0} $ & $23.2 \pm 1.4$ \\
  $\gamma^{[deg]} $ & $ 62^{+10}_{-12}$ & $61.6 \pm 7 $ \\
  \hline
\end{tabular}

\section{The value of $B_K$}\label{extraction}

From the results of the fit, presented above, we can extract the
value of $B_K$. For this purpose we use the theoretical expression
for $|\tilde{\epsilon}|$, first obtained in \cite{Vysold}. It has
the following form:
\begin{eqnarray}\label{tilde epsilon}
  &&|\tilde{\epsilon}^{theo}|=\frac{G^2_F m_K f^2_K}{12\sqrt{2}\pi^2\Delta m_K}
  B_K (\eta_{cc} m_c^2 Im[
  (V_{cs} V^*_{cd})^2]+\eta_{tt} m_t^2 I(\xi) Im[(V_{ts} V^*_{td})^2]\nonumber\\
  &&+2\eta_{ct} m_c^2 \ln (\frac{m_W^2}{m_c^2})  Im[V_{cs} V^*_{cd} V_{ts} V^*_{td}] ).
\end{eqnarray}

Here $\displaystyle{I(\xi) = \{\frac{\xi^2-11 \xi + 4}{4
(\xi-1)^2}- \frac{3 \xi^2 \ln \xi}{2 (1-\xi)^3}\}}$,
$\xi=m^2_t/m^2_W$. Quark masses are $m_c=1.2\pm 0.2$ GeV
\cite{PDG}, $m_t=178.0\pm 4.3$ GeV \cite{mt}, $m_W=80.42\pm 0.04$
GeV \cite{PDG}. The QCD corrections were calculated to leading
order in \cite{Vysold}: $\eta_{cc}=0.6$, $\eta_{tt}=0.6$,
$\eta_{ct}=0.4$. The next-to-leading order calculation changes
slightly $\eta_{tt}$ and $\eta_{ct}$ and changes considerably
$\eta_{cc}$: $\eta_{cc}=1.32\pm 0.32$ \cite{her}, $\eta_{tt}=0.574
\pm 0.01$ \cite{burasj}, $\eta_{ct}=0.47 \pm 0.04$ \cite{herr}.
The kaon decay constant extracted from the $K^+\rightarrow \mu^+
\nu$ decay width equals: $f_K=160.4\pm 1.9$ MeV \cite{PDG}. The
$K_L-K_S$ mass difference is $\Delta m_K=(3.483\pm 0.006)\times
10^{-15}$ GeV \cite{PDG}. Fermi constant $G_F=1.16639(1)\times
10^{-5} GeV^{-2}$\cite{PDG}.

Now we equate this expression to the value of $|\tilde{\epsilon}|
= (2.51 \pm 0.07) \times 10^{-3}$ from Eq.(\ref{10}), substituting
all experimental numbers and the results of the fit. This leads to
the following value of $B_K$:

\begin{equation}
B_K = 0.89 \pm 0.16
\end{equation}

Note that it is close to the result of vacuum insertion: $B_K = 1$.

%
%

\section{Conclusions}\label{concl}

We have extracted the value of $B_K$ using the fitted values of
CKM matrix elements and the estimated difference between $\tilde
\epsilon$ and $\epsilon$. Our result is $B_K = 0.89 \pm 0.16$. It
appears to be  close to the result of vacuum insertion, $B_K = 1$,
while lattice result is simply the same: $B_K = 0.87 \pm 0.06 \pm
0.14_{quench}$ \cite{Lel}.

\section*{Acknowledgements}

We are grateful to Augusto Ceccucci and Ed Blucher for providing
us with the latest experimental data on $\delta_L$. This work was
partially supported by the program 
FS NTP FYaF 40.052.1.1.1112 and by 
grant NSh- 2328.2003.2. G.O. is grateful to Dynasty Foundation for
partial support.

\appendix
\section{Estimation of the value of $Im A_0$ from
QCD penguin diagram.}\label{penguin}

Let's estimate $\frac{Im A_0}{Re A_0}$, using experimental value
of $Re A_0$ and evaluating the value of $Im A_0$. The latter will
be evaluated to the following accuracy: the LO Wilson coefficients
will be used, and hadronic matrix element will be calculated in
naive factorization approximation.

As it is well known transitions with $\Delta S = 1$ are due to the
4-quark effective Hamiltonian, for the first time derived in
\cite{Shifman}:

\begin{eqnarray}
H_{\Delta S=1} = \sqrt{2} G_F \sin \theta_C \cos \theta_C
\sum_{i=1}^{6} c_i O_i
\end{eqnarray}

The so-called penguin operator $O_5$ dominates in amplitudes $K^0
\rightarrow (\pi \pi)_{I=0}$ \cite{Shifman}:

\begin{equation}
    O_5 = \bar s_L \gamma_{\mu} \lambda^a d_L (\bar u_R \gamma_{\mu} \lambda^a u_R +
    \bar d_R \gamma_{\mu} \lambda^a d_R)
\end{equation}

Below we present the detailed derivation of the coefficient function $c_5$
in one loop approximation.

\begin{figure}[!htb]
\centering \epsfig{file=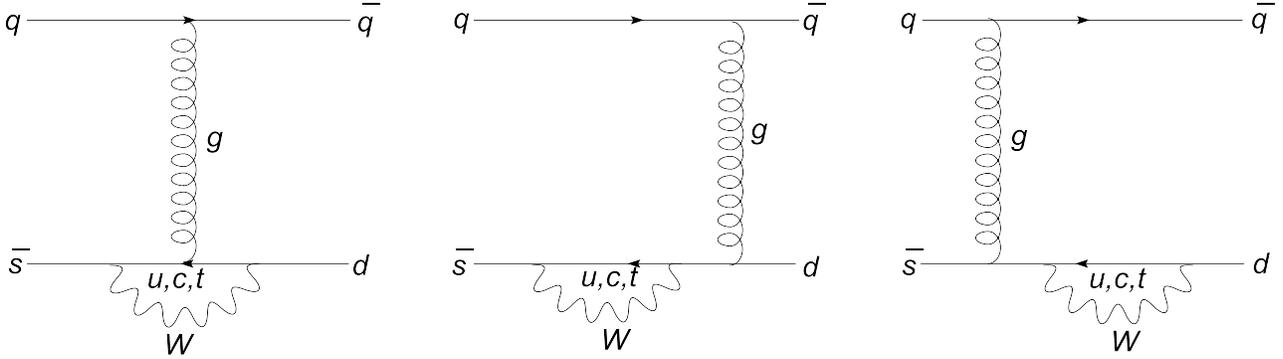,width=17cm,height=5cm}
\caption{\em  Diagrams, which contribute to the penguin operator.}
\label{penguins}
\end{figure}

It is convenient to perform calculation in the unitary gauge. Each
of the three diagrams (see Fig.~\ref{penguins}) is infinite,
however the sum appears to be finite. We will use the dimensional
regularization ($d=4-2\epsilon$), in order to regularize divergent
integrals:

\begin{eqnarray}\label{diagrams}
  &&M_0=\frac{g^2 g_s^2}{2} \frac{1}{q^2} \sum_{j=u,c,t} \int\frac{d^d k}{(2\pi)^d}\frac{g_{\mu \nu}-\frac{k_\mu k_\nu}{M_W^2}}
{k^2-M_W^2} \bar{s}_L\gamma_{\mu}\frac{1}{\hat{q}_2-m_j}
\gamma_\rho \frac{\lambda_a}{2}\frac{1}{\hat{q}_1-m_j}\gamma_\nu d_L \left(\bar{\psi}_r \gamma_\rho \frac{\lambda_a}{2}\psi_r\right)V_{jd} V_{js}^*,\nonumber\\
  &&M_1=\frac{g^2 g_s^2}{2} \frac{1}{q^2} \sum_{j=u,c,t} \int\frac{d^d k}{(2\pi)^d}\frac{g_{\mu \nu}-\frac{k_\mu k_\nu}{M_W^2}}
{k^2-M_W^2} \bar{s}\gamma_{\mu}p_L\frac{1}{\hat{q}_2-m_j}
\gamma_\nu p_L  \frac{1}{\hat{p}_2-m_j} \gamma_\rho
\frac{\lambda_a}{2}
d   \left(\bar{\psi}_r \gamma_\rho \frac{\lambda_a}{2}\psi_r\right) V_{jd} V_{js}^*,\nonumber\\
  &&M_2=\frac{g^2 g_s^2}{2} \frac{1}{q^2} \sum_{j=u,c,t} \int\frac{d^d k}{(2\pi)^d}\frac{g_{\mu \nu}-\frac{k_\mu k_\nu}{M_W^2}}
{k^2-M_W^2} \bar{s}\gamma_\rho \frac{\lambda_a}{2}
\frac{1}{\hat{p}_1-m_j}\gamma_{\mu}p_L \frac{1}{\hat{q}_1-m_j} \gamma_\nu p_L d \left(\bar{\psi}_r\gamma_\rho \frac{\lambda_a}{2} \psi_r\right) V_{jd} V_{js}^*,\nonumber\\
  && M=M_0+M_1+M_2.
\end{eqnarray}

The effective Hamiltonian is equal to:

\begin{equation}\label{O_def}
  H_{penguin}=i M.
\end{equation}

The result of the calculation can be presented in the following
form:

\begin{eqnarray}\label{O_result}
  &&H_{penguin}=-\sum_{j=u,c,t}\frac{g^2 g_s^2}{48\pi^2} \frac{1}{q^2}\bar{s}_L G_\rho(q) \frac{\lambda_a}{2} d_L
\left(\bar{\psi}_r \gamma_\rho \frac{\lambda_a}{2}\psi_r\right)V_{jd} V_{js}^*,\nonumber\\
  &&G_\rho=G_1\gamma_\rho+\frac{1}{M_W^2}\left(G_2 \hat{p}_2\gamma_\rho\hat{p}_1+G_3\hat{p}_1 p_{1\rho}+G_4\hat{p}_2 p_{2\rho}
+G_5\hat{p}_1 p_{2\rho}+G_6\hat{p}_2 p_{1\rho}\right).
\end{eqnarray}

Dimensionless functions $G_i$ ($i=1,2,3,4,5,6$) depend on the
values $x_j=\frac{m_j^2}{M_W^2}$, $q^2$, $m_s^2$, $m_d^2$, where
$m_j$ is the up-quark mass in the loop. We suppose the external
$s$ and $d$ quarks to be on mass shell.

Let us neglect d-quark mass. In this approximation the
non-zero contribution into operator $H_{penguin}$ is given by the
terms with formfactors $G_1, G_4$ and $G_6$. It is convenient to
introduce new variables $P=p_1+p_2$ and $q=p_1-p_2$:

\begin{equation}\label{G_simp}
\bar{s}_L G_\rho \frac{\lambda_a}{2}d_L=G_1
\bar{s}_L\gamma_\rho\frac{\lambda_a}{2}d_L+\frac{m_s
\left((G_6+G_4)P_{\rho}+(G_6-G_4)q_{\rho}\right)}
{2M_W^2}\bar{s}_R \frac{\lambda_a}{2}d_L.
\end{equation}

Here the term proportional to $q_\rho$ will not contribute to
operator $H_{penguin}$, since $q_\rho\times \left(\bar{\psi}_r
\gamma_\rho \frac{\lambda_a}{2}\psi_r\right) = (m_{r}-m_{r})
\left(\bar{\psi}_2 \frac{\lambda_a}{2}\psi_1\right)=0$. The
quantity $P_\rho\bar{s}_R\frac{\lambda_a}{2}d_L$ should be
expressed through the magnetic formfactor with the help of the
following equation:

\begin{equation}\label{24}
  \bar{s}\sigma_{\mu \nu} q_\nu \frac{\lambda_a}{2}d_L=\frac{i}{2}q_\nu\bar{s}
  (\gamma_\mu\gamma_\nu-\gamma_\nu\gamma_\mu)
\frac{\lambda_a}{2}d_L =i(m_s\bar{s}_L\gamma_\mu
\frac{\lambda_a}{2}d_L-P_\mu \bar{s}_R\frac{\lambda_a}{2}d_L).
\end{equation}

With the help of Eq.(\ref{24}) from Eq.(\ref{G_simp}) we obtain:

\begin{equation}\label{4quark_operat}
  H_{penguin}=-\sum_{j=u,c,t}\frac{g^2 g_s^2}{48 \pi^2} \frac{1}{q^2}\left(f_1 \bar{s}_L\gamma_\mu\frac{\lambda_a}{2}d_L+i \frac{f_2}{M_W^2}
m_s q_{\nu}\bar{s}_R \sigma_{\mu\nu}\frac{\lambda_a}{2}d_L\right)
\bar{\psi}_r \gamma_\mu \frac{\lambda_a}{2}\psi_r V_{jd} V_{js}^*,
\end{equation}
where $f_1$ and $f_2$ are equal to:
\begin{eqnarray}\label {f1f2}
  &&f_1=G_1+\frac{m_s^2}{2M_W^2}(G_6+G_4),\nonumber\\
  &&f_2=\frac{G_6+G_4}{2}.
\end{eqnarray}

It is sufficient to calculate the formfactors $G_4$ and $G_6$ in
the zero order in $q^2$, $m_s^2$. However, the formfactor $G_1$,
as it follows from last equations, should be calculated, including
terms proportional to $m_s^2$, $q^2$. From equations
(\ref{diagrams}), calculating appropriate integrals, we get:

\begin{eqnarray}
  &&G_1=R_1\frac{q^2}{M_W^2}+R_2\frac{m_s^2}{M_W^2},\nonumber\\
  &&R_1=\frac{7x^4+14x^3-63x^2+38x+4+6(16x-9x^2-4)\ln x}{24(1-x)^4},\nonumber\\
  &&R_2=\frac{-5x^4+14x^3-39x^2+38x-8+18x^2 \ln x}{8(1-x)^4},\nonumber\\
  &&G_4=\frac{2x^4-14x^3+45x^2-38x+5+6(1-4x)\ln x}{6(1-x)^4},\nonumber\\
  &&G_6=\frac{11x^4-14x^3+27x^2-38x+14+6(8x-9x^2-2)\ln x}{12(1-x)^4}.
\end{eqnarray}

Substituting these formulas into equations (\ref{f1f2}), we
obtain:
\begin{eqnarray}\label {f1f2_result}
  &&f_1=\frac{7x^4+14x^3-63x^2+38x+4+6(16x-9x^2-4)\ln x}{24(1-x)^4}\frac{q^2}{M_W^2},\nonumber\\
  &&f_2=\frac{5x^4-14x^3+39x^2-38x+8-18x^2 \ln x}{8(1-x)^4}.
\end{eqnarray}

Finally, let us rewrite equation (\ref{4quark_operat}) in the
following way:
\begin{eqnarray}\label{final}
    &&H_{penguin}=-\sum_{j=u,c,t}\frac{g^2 g_s^2}{48 \pi^2 M_W^2}\left(F_1 \bar{s}_L\gamma_\mu\frac{\lambda_a}{2}d_L+i
F_2 m_s \frac{q_{\nu}}{q^2}\bar{s}_R
\sigma_{\mu\nu}\frac{\lambda_a}{2}d_L\right)
\bar{\psi}_r \gamma_\mu \frac{\lambda_a}{2}\psi_r V_{jd} V_{js}^*,\nonumber\\
    &&F_1=\frac{7x^4+14x^3-63x^2+38x+4+6(16x-9x^2-4)\ln x}{24(1-x)^4},\nonumber\\
    &&F_2=f_2=\frac{5x^4-14x^3+39x^2-38x+8-18x^2 \ln x}{8(1-x)^4}.
\end{eqnarray}

As the admixture of gluons in $K$ and $\pi$ mesons is small, the
contribution of magnetic moment operator in (\ref{final}) is
negligible \cite{Shifman}.

Substituting $m_c=1.2$ GeV, $m_t=178.0$ GeV, $M_W=80.42$ GeV for
the formfactor $F_1$ we obtain:
\begin{eqnarray}\label{aprox}
  &&F_1(x<<1)\approx -\ln x +\frac{1}{6}  \nonumber,\\
  &&F_1(x_c)\approx 8.58\nonumber,\\
  &&F_1(x_t)\approx 0.550\nonumber,\\
  &&F_1(\infty)=\frac{7}{24}\approx 0.292.
\end{eqnarray}

Formula (\ref{final}) can be rewritten with good accuracy as:
\begin{eqnarray}\label{13}
H_{penguin}\approx\sqrt{2}G_F\sin{\theta_C}\cos{\theta_C}\left(-\frac{\alpha_s}{12\pi}
\ln\frac{m_c^2}{\mu^2}+ i \frac{Im V_{cd} V_{cs}^*}{Re V_{cd}
V_{cs}^*} \frac{\alpha_s}{12\pi} \ln\frac{M_W^2}{m_c^2} \right)
O_5,
\end{eqnarray}

\noindent where instead of $m_u$ the characteristic hadronic scale
$\mu$ (this time ``low'' normalization point) is substituted.

Thus the real and imaginary parts of $c_5$ are equal to:

\begin{eqnarray}\label{14}
&& Re c_5 = -\frac{\alpha_s}{12\pi} \ln\frac{m_c^2}{\mu^2}\nonumber\\
 &&Im c_5 = \frac{Im V_{cd}
V_{cs}^*}{Re V_{cd} V_{cs}^*}\frac{\alpha_s}{12\pi}
\ln\frac{M_W^2}{m_c^2}.
\end{eqnarray}

In order to understand at which virtuality $\alpha_s$ should be
taken in these expressions leading
logarithms should be summed up. This was done for
the real part of coefficient function in the paper \cite{Shifman}:

\begin{eqnarray}
&&Re c_5 = \left( \chi_1^{0.48} \left(- 0.039\,\chi_2^{ 0.8}+
0.033\,\chi_2^{ 0.42}+ 0.003 \,\chi_2^{-
0.12}+ 0.003\,\chi_2^{- 0.3} \right)+ \right.\nonumber\\
&&\left.+\chi_1^{-0.24} \left(- 0.014\,\chi_2^{ 0.8}-
0.001\,\chi_2^{ 0.42}- 0.014 \,\chi_2^{- 0.12}+ 0.029\,\chi_2^{-
0.3} \right)\right),
\end{eqnarray}

\noindent while for imaginary part in paper \cite{Guberina} the
following result was obtained:

\begin{eqnarray}
&&Im c_5 = \frac{Im V_{cd} V_{cs}^*}{Re V_{cd} V_{cs}^*} \left(
0.0494 \,\chi_1^{ 0.85}- 0.0280\,\chi_1^{ 0.42}+
 0.0116\,\chi_1^{- 0.13}- 0.0330\,\chi_1^{- 0.35} \right) \times
\nonumber\\
&&\times \left(  0.8509\,\chi_2^{ 0.8}+ 0.0091\,\chi_2^{ 0.42}+
0.1222 \,\chi_2^{- 0.12}+ 0.0178\,\chi_2^{- 0.3} \right),
\end{eqnarray}

\noindent where $\chi_1 = \frac{\alpha_s(m_c)}{\alpha_s(m_W)}$,
$\chi_2 = \frac{\alpha_s(\mu)}{\alpha_s(m_c)}$.

\vphantom{12pt}

Numerical analysis shows that with a good accuracy expression for
$Re c_5$ can be written as:

\begin{eqnarray}\label{112}
&&Re c_5 = -\frac{\alpha_s(\mu)}{12 \pi} \log
(\frac{m_c^2}{\mu^2}).
\end{eqnarray}

On the other hand $Im c_5$ at the scale $\mu$ at which
$\alpha_s(\mu) = 1$ has the value

\begin{equation}
Im c_5 = \frac{Im V_{cd} V_{cs}^*}{Re V_{cd} V_{cs}^*} \times
0.13.
\end{equation}

The expression for $Im A_0$ can be written as:

\begin{eqnarray}
Im A_0 = \sqrt{2} G_F \sin \theta_C \cos \theta_C Im(c_5) <(\pi
\pi)_{I=0}|O_5|K^0>
\end{eqnarray}

In order to get the value of $Im A_0$ we must calculate hadronic
matrix element of penguin operator. It was evaluated in the framework
of naive quark model in \cite{Shifman}, see also \cite{Vainstein}:
\begin{eqnarray}
<(\pi \pi)_{I=0}|O_5|K^0> = \frac{4 \sqrt{6}}{9 } \frac{m_K^2
m_{\pi}^2 f_{\pi}}{m_s (m_u+m_d)} \left( \frac{f_K}{f_{\pi}}
(1+\frac{m_K^2}{m_{\sigma}^2}) - 1\right)
\end{eqnarray}

Substituting experimental numbers $G_F = 1.166 \times 10^{-5} GeV,
\sin \theta_C =0.22, \cos \theta_C = 0.95, m_{\pi}=135 MeV,
m_K=497 MeV, f_{\pi}=130 MeV ,f_K=160 MeV ,m_{\sigma}=700 MeV$ and
quark masses $m_s=130 MeV, m_u=3 MeV, m_d=7 MeV$, we get:

\begin{eqnarray}
Im A_0 = -1.1 \times 10^{-10} GeV.
\end{eqnarray}

Finally, dividing it by experimentally measured $Re A_0 = 3.33
\times 10^{-7} GeV$, we obtain:

\begin{eqnarray}\label{115}
\frac{Im A_0}{Re A_0} = - 3.2 \times 10^{-4}.
\end{eqnarray}

In order to estimate the theoretical error for $Im A_0$ we
propose the following method: to take two values of $\mu$,
corresponding to $\alpha_s(\mu) = \frac{2}{3}$ and $\alpha_s(\mu)
= \frac{3}{2}$, to calculate $\frac{Im A_0}{Re A_0}$ at each $\mu$ and
from these boundary values get $\pm$ error for $\frac{Im A_0}{Re
A_0}$.

Via the proposed method we get our final result:

\begin{eqnarray}\label{116}
\frac{Im A_0}{Re A_0} = -( 3.2^{+1.1}_{-0.8} )\times 10^{-4}.
\end{eqnarray}

This number is
rather stable with respect to the variation
of $\mu$. Our result confirms statement maid in \cite{Vainstein}:
QCD penguin results in the value of $\displaystyle{\frac{\epsilon'}{\epsilon}}$
in ballpark of the experimental data.

On the other hand the real part is very sensitive to $\mu$, as can
be seen from Eq.(\ref{112}). The expression for $Re A_0$ has the
following form:

\begin{eqnarray}
Re A_0 = \sqrt{2} G_F \sin \theta_C \cos \theta_C Re(c_5) <(\pi
\pi)_{I=0}|O_5|K^0>.
\end{eqnarray}

Substituting numbers and again taking $\mu$ at which
 $\alpha_s(\mu) = \frac{2}{3}$ and $\alpha_s(\mu)
= \frac{3}{2}$ we get: $Re A_0 = (1.2^{+0.8}_{-0.6}) \times
10^{-7} GeV$. The central number is approximately 3 times smaller than
the experimental , but theoretical uncertainty is large.

\end{document}